\documentclass[a4paper]{jpconf}
\usepackage{graphicx}
\def\beq{\begin{equation}}
\def\eeq{\end{equation}}
\def\bea{\begin{eqnarray}}
\def\eea{\end{eqnarray}}
\def\bq{\begin{quote}}
\def\eq{\end{quote}}

\newcommand{\gsim}{\lower.7ex\hbox{$\;\stackrel{\textstyle>}{\sim}\;$}}
\newcommand{\lsimm}{\lower.7ex\hbox{$\;\stackrel{\textstyle<}{\sim}\;$}}
\begin{document}
\title{Particle Dark Matter Candidates}

\author{Stefano Scopel}

\address{
Korea Institute for Advanced Study\\
207-43 Cheongryangri-dong Dongdaemun-gu\\
Seoul 130-722, Korea}

\ead{scopel@kias.re.kr}

\begin{abstract}
I give a short overview on some of the favorite particle Cold Dark
Matter candidates today, focusing on those having detectable
interactions: the axion, the KK--photon in Universal Extra Dimensions,
the heavy photon in Little Higgs and the neutralino in Supersymmetry.
The neutralino is still the most popular, and today is available in
different flavours: SUGRA, nuSUGRA, sub--GUT, Mirage mediation, NMSSM,
effective MSSM, scenarios with CP violation.  Some of these scenarios
are already at the level of present sensitivities for direct DM
searches.
\end{abstract}

\section{Introduction}

About 20\% of the energy density of the present Universe appears to be
in the form of Cold Dark Matter (CMD) \cite{wmap}, i.e. of particles
non--relativistic at the time of their decoupling from the plasma in
the Early Universe whose contribution is considered as an
indispensable catalyst to the formation of galaxies. CDM clusters at
the galactic level, so that it should pervade our solar system and be
detectable by direct or indirect techniques. In my talk I will give a
short overview on some of the most popular particle CDM candidates
today, focusing on those having detectable interactions. So, while
Super--Weak DM candidates such as sterile neutrinos and gravitinos
represent a viable possibility to explain DM \cite{superweak}, I will
not cover them here since they cannot be measured in DM
searches. Moreover, I will only discuss direct DM detection, since
indirect searches have been discussed elsewhere in this Conference
\cite{indirect}.

The properties of a good DM candidate are well known: it should be
stable (its decay being forbidden by the conservation of some quantum
number), neutrally charged and colourless (in order to be really
``dark''), possibly motivated by theory (although ``ad hoc''
candidates may have the advantage of minimality and simplicity
\cite{ad_hoc}). Moreover, its calculated relic abundance should be
compatible to observation, although candidates providing a subdominant
contribution to CDM represent a quite reasonable
option\footnote{Variety is common in Nature, and a multicomponent CDM
might even have larger detection cross sections, and so be easier to
detect.}.

In the following $\Omega\equiv \rho/\rho_c$ will indicate a candidate
cosmological mass density $\rho$ normalized to the critical one,
$\rho_c=$1.8791 h$^{2}\times$ 10$^{-29}$ g cm$^{-3}$, and $h\equiv
H_0/100$ km sec$^{-1}$ Mpc$^{-1}$ will indicate the normalized Hubble
constant at present times.

\section{The neutrino}

\begin{figure}[t]
\begin{center}
\hspace{-1cm}
\includegraphics*[height=7cm]{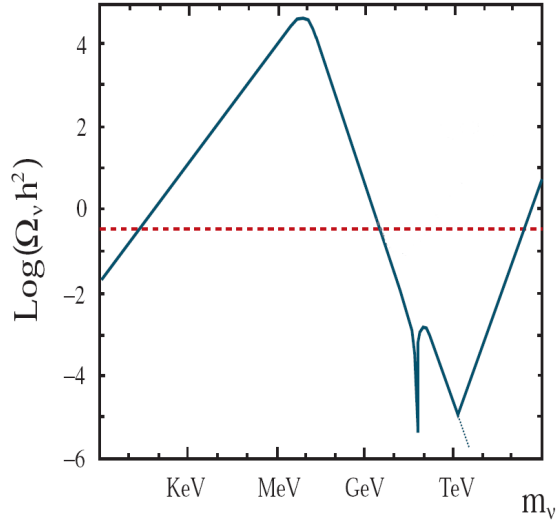}
\end{center}
\caption{Neutrino relic abundance as a function of the neutrino mass
(from\protect\cite{neutrino})
\label{fig:neutrino_relic_abundance}}
\end{figure}

The first place to look for a DM candidate is the Standard Model (SM)
of interactions. In this case the only viable DM candidate is the
neutrino, whose relic abundance as a function of the mass is shown in
Fig. \ref{fig:neutrino_relic_abundance}, taken from
\cite{neutrino}. Since LEP excludes additional neutrinos (with
standard weak interactions) with masses $m_{\nu}\lsimm$ 45 GeV one can
see from this figure that the mass density of ordinary heavy neutrinos
is bound to be very small, $\Omega_{\nu} h^2 \lsimm$ 0.001 for masses
up to $m_{\nu}\sim{\cal O} (100)$ TeV.  At lower masses the neutrino
relic abundance is given by $\Omega_{\nu} h^2=\sum m_{\nu}/91.5$ eV,
that, when compared to observation, implies the constraint $\sum
m_{\nu}\lsimm$ 10 eV. However a more stringent limit $\sum
m_{\nu}\lsimm$ 0.66 eV can be derived combining CMB and Large Scale
Structure (LSS) data \cite{wmap}, implying $\Omega_{\nu} h^2\lsimm
0.007$: in fact, neutrinos lighter than this bound, which are
relativistic ({\it hot}) at decoupling, erase primordial perturbations
due to their free streaming and prevent galaxy formation. So also in
this case the neutrino abundance turns out to be small. The bottom
line is that, unless neutrinos are mixed with some sterile component
\cite{superweak} they don't work as DM candidates.

The need of a DM candidate outside the SM, when combined with the fact
that the SM itself is believed to be an incomplete theory with a
cut--off of ${\cal O}\sim$ 1 TeV, is no doubt intriguing. Moreover,
theory is in no shortage of viable DM candidates that have been proposed in
the context of Particle Physics in order to solve/alleviate/explain
problems that have nothing to do with Cosmology.  In the following I
will concentrate on some of the most popular.

\section{The axion}
\begin{figure}[t]
\begin{center}
\hspace{-1cm}
\includegraphics*[height=7cm]{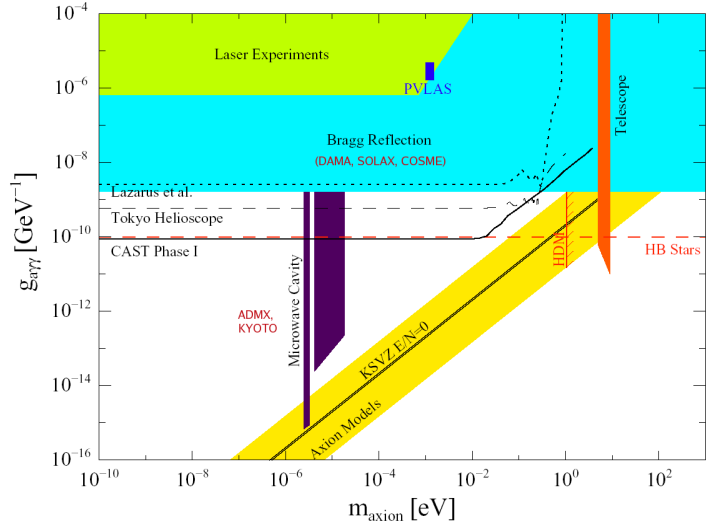}
\end{center}
\caption{Constraints on the axion--photon coupling as a function of
the axion mass. 
\label{fig:axion_limits}}
\end{figure}

The axion is the pseudo Goldstone boson of the Peccei--Quinn (PQ)
symmetry, introduced in order to explain CP conservation in QCD
\cite{axion}. Its mass can be considered as a free parameter, spanning
several orders of magnitude, and is related to the (unknown) scale of
the PQ symmetry breaking $f_a$. The main production mechanism of relic
axions is through misalignment, in which at $T>\Lambda_{QCD}$ the
axion field acquires a random initial phase $\theta_i$ in its flat
potential, while later on, when $T<\Lambda_{QCD}$ and the potential
gets ``tilted'', it starts to oscillate coherently around the minimum,
behaving like pressureless non--relativistic DM with density $\Omega_a
h^2\simeq k_a (m_a/10^{-5} {\rm eV})^{-1.175} \theta_i^2$ with
$0.3<k_a$ a few. If one assumes no inflation after the PQ phase
transition, $\theta_i$ is randomly distributed in the Universe, so
that making the average, $<\theta_i^2>=\pi^2/3$, the relic abundance
falls in the observed range for $m_a\gsim 10^{-5}$ eV. Axions can be
observed through their conversion to photons when crossing a magnetic
field. The present experimental situation is shown in
Fig.\ref{fig:axion_limits}, where axion models span the yellow
(oblique) band. Note however that the axion--photon coupling is
affected by uncertainties on the light quark masses so that it might
be suppressed compared to what usually
expected\cite{axion_uncertainties}. Axions are also produced today in
the interior of the Sun and may be measured by making use of resonant
Bragg reflection in the same crystal detectors used for direct DM
searches\cite{solar_axion}. However their sensitivity scales as the
exposition to the power 1/8, so that present limits have already
saturated their maximal reach and cannot be significantly improved.

\section{Weakly Interacting Massive Particles (WIMPs)}
The appeal of WIMPs as DM candidates lies in the fact that their
thermal cosmological density, given by: $\Omega_{WIMP}\simeq 10^{-38}
{\rm cm^3 sec^{-1}}/<\sigma_{ann} v>_{int}$, falls naturally in the
correct range if $<\sigma v>_{int}$ is of the order of weak--type
interactions. In the last expression $<\sigma v>_{int}$ is the
integral over T from the WIMP decoupling temperature $T_f$ to the
present one of the thermal average of the WIMP annihilation cross
section $\sigma_{ann}$ to standard particles. This simple picture is
modified when the WIMP is close in mass to other particles that may
either co--annihilate faster, depleting its density, or slower,
enhancing the density by acting as a WIMP reservoir at their decay.
Moreover, the cross section may be strongly enhanced (and the relic
density suppressed) for particular values of the WIMP mass for which
resonant annihilation takes place. Examples of WIMPs are the Heavy
Photon in Little Higgs theories, the KK--photon in Universal Extra
Dimensions and the neutralino in Supersymmetry.

In the following $\sigma^{\rm (nucleon)}_{\rm scalar}$ will indicate
the coherent WIMP--nucleon cross section.

\subsection{Little Higgs (LH)}

In LH models \cite{little_higgs} the Higgs particle is naturally light
because it is the pseudo Goldstone boson from the (collective)
symmetry breaking of an effective non--linear $\sigma$ model. New
heavy particles (gauge bosons, singlet and triplet scalars, neutrinos,
quarks) are present in the model, and T--parity conservation (where SM
particles are even and new ones mostly odd) is introduced in order to
forbid dangerous tree level couplings that would put severe
constraints from precision Electro--Weak (EW) tests\footnote{Note
however that it has been pointed out that T--parity is generally
broken by anomalies\protect\cite{t_parity_violation}}. The Lighest
T--odd Particle (LTP) is typically a Heavy Photon $B_H$, which is
stable and can be a dark matter candidate. Having weak--type
interactions the relic abundance of $B_H$ falls naturally in the
correct range. On the other hand, its direct detection cross section
is expected to be smaller than the present experimental sensitivities
(barring some resonant effect), $\sigma^{\rm (nucleon)}_{\rm
scalar}\lsimm 10^{-9}$ pbarn.

\subsection{Universal Extra Dimensions (UED)}

In the simplest UED model\cite{ued} all SM fields propagate in an
additional 5$^{th}$ dimension which is compactified on an orbifold
$S_1/Z_2$. The dispersion relation in the 5$^{th}$ dimension,
$E^2=\vec{p}^2+(p_5^2+M^2)$ implies for each SM particle an infinite
tower of massive states (KK tower) in the effective 4--dimensional
theory, with $p_5=n/R$ ($n=1,2,3,...)$ while, from electroweak tests
constraints, $R^{-1}\gsim 300$ GeV. Orbifold compactification,
required in order to get rid of unwanted degrees of freedom in the
ground (n=0) states, breaks translational invariance in the 5$^{th}$
dimension, but leaves unbroken the residual invariance under discrete
$\pi R$ translations (KK--parity$\equiv(-1)^n$).  Typically, the
Lighest KK particle is the KK photon $B^{(1)}$ which is stable and a
DM candidate. Because of the peculiar evenly--spaced spectrum of KK
particles, many modes in UED have similar masses (implying
co--annihilations) or integer mass ratios (implying resonant
annihilations). In particular, depending on whether the
co--annihilating next-to-lighest KK particles are strongly interacting
(KK quarks and/or gluons) or not (KK leptons) the relic abundance may
be either suppressed or enhanced, so that the range of $R^{-1}$ which
corresponds to a correct DM abundance is rather large: 500 GeV$\lsimm
R^{-1}\lsimm$ 2 TeV. Direct detection turns out to be below present
sensitivities, $\sigma^{\rm (nucleon)}_{\rm scalar}\lsimm 10^{-9}$
pbarn, and depends on the degree of degeneracy between $B^{(1)}$ and
the KK quarks in the propagator.

\subsection{Supersymmetry (susy)}

Susy is widely considered to be the most natural extension of the
SM. In susy every SM particle belongs to a supermultiplet
containing partners with opposite statistics (the susy partners of
gauge and Higgs bosons being fermionic gauginos and Higgsinos,
while those of fermions being scalar fermions or sfermions) in
such a way that the total number of fermionic and bosonic degrees
of freedom is the same. In general the theory has Yukawa couplings
involving squarks and sleptons that violate the baryon and lepton
numbers at the tree level and that imply fast proton decay.
R--parity is the symmetry that the theory acquires when all these
couplings are removed from the superpotential: all SM particles
are parity--even, while susy partners are parity--odd, so if
R--parity is conserved the Lighest Supersymmetric Particle (LSP)
is stable and can be a DM candidate. Moreover, since we know from
experiment that particles within supermultiplets are not
degenerate in mass, susy must be broken. Unfortunately the
mechanism of susy-breaking is model--dependent, so the theory has
many free parameters: soft and gaugino masses, the Higgs--mixing
parameter $\mu$ and the ratio of the two vacuum expectation values
$\tan\beta$ (in susy at least two Higgs doublets are required, and
the particle content, consisting in 5 states, contains 2 neutral
scalars $h$ and $H$, one neutral pseudoscalar $m_A$ and 2 charged
scalars $H^{\pm}$) along with a huge number of flavour--mixing
parameters and phases that are usually neglected. In a large
region of the susy parameter space the LSP is the neutralino
\footnote{For a recent reassessment of sneutrino DM, see
\protect\cite{arina}}, defined as the superposition of neutral
gauginos and Higgsinos: $ \chi\equiv a_1 \tilde{B}+a_2 \tilde{W_3}+a_3
\tilde{H}_1 +a_4 \tilde{H}_2$, which clearly has all the properties of
an ideal DM candidate. In the following I will discuss the neutralino
DM phenomenology in three possible scenarios: SuperGRAvity inspired
(SUGRA), Next-to-Minimal SuSy Model (NMSSM) and effective
MSSM. Non--Universal SUGRA (nuSUGRA)\cite{nusugra}, sub-GUT
\cite{sub-GUT}, Mirage mediation \cite{mirage_mediation}, MSSM singlet
extensions \cite{langacker} and CP violating models \cite{susy_cpv}
are other scenarios that can be found in the literature and that I
will not cover here.

The SUGRA scenario\cite{SUGRA} has just 5 free parameters, all defined
at the Grand Unification (GUT) scale: one common soft mass $m_0$ for
scalars, one common soft mass $m_{1/2}$ for gauginos, $\tan\beta$, the
trilinear coupling $A_0$ and the sign of $\mu$. They are evolved down
to the EW scale by making use of renormalization group
equations. In this scenario EW symmetry breaking is achieved
radiatively, since the large top Yukawa coupling drives one Higgs mass
parameter negative in the running from the GUT to the EW scale. This
typically implies a large $\mu$ parameter with $\chi\simeq \tilde{B}$
and a large value for $m_A$ unless $\tan\beta\simeq$ 50. When present
experimental limits are taken into account only particular corners of
the SUGRA parameter space turn out to be compatible to observation,
because the neutralino relic abundance is typically too high. The
cosmologically allowed regions correspond to the so--called stau
co-annihilation, Higgs funnel and focus point scenarios
\cite{SUGRA}. The corresponding direct detection cross section is
marginally at the level of present sensitivities, $\sigma^{\rm
(nucleon)}_{\rm scalar}\lsimm 10^{-7}$ pbarn
\cite{sugra_ellis_diretta}.

\begin{figure}
\begin{center}
\hspace{-1cm} \includegraphics*[height=7cm]{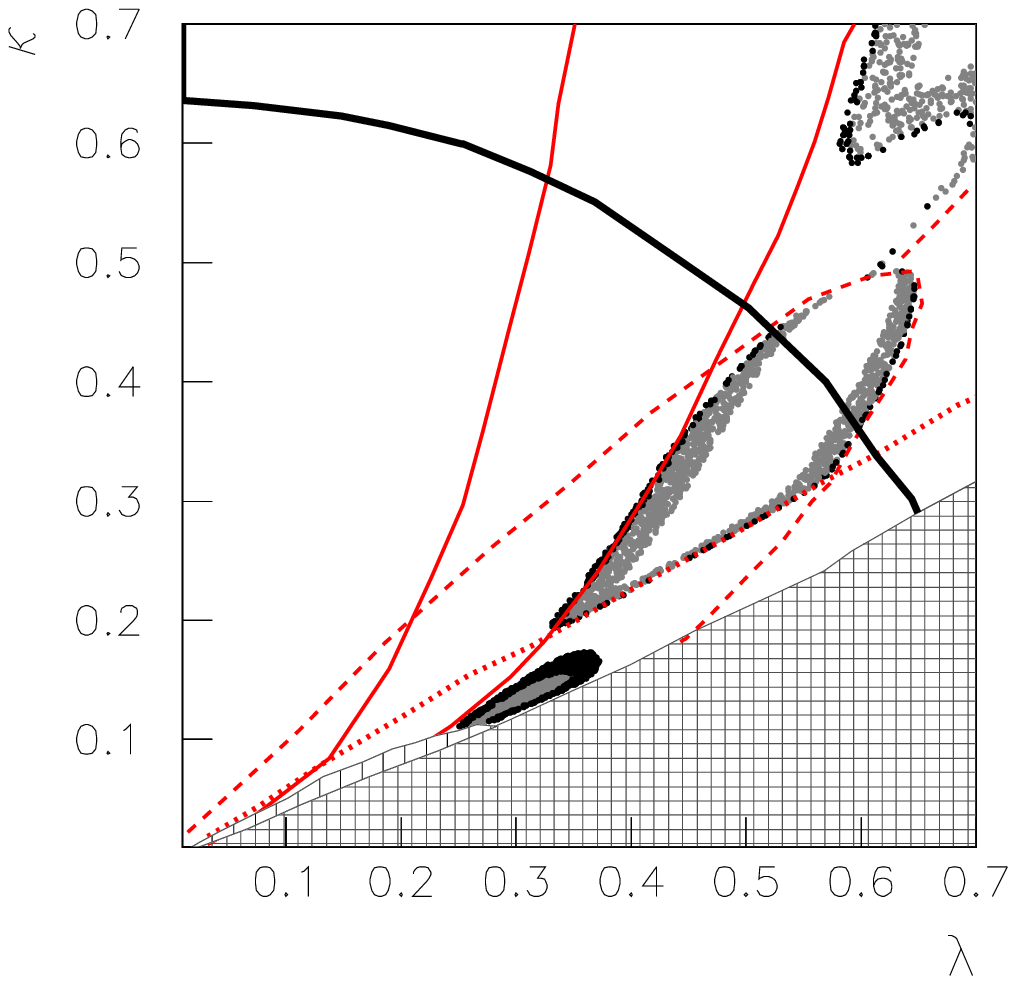}
\includegraphics*[height=7cm]{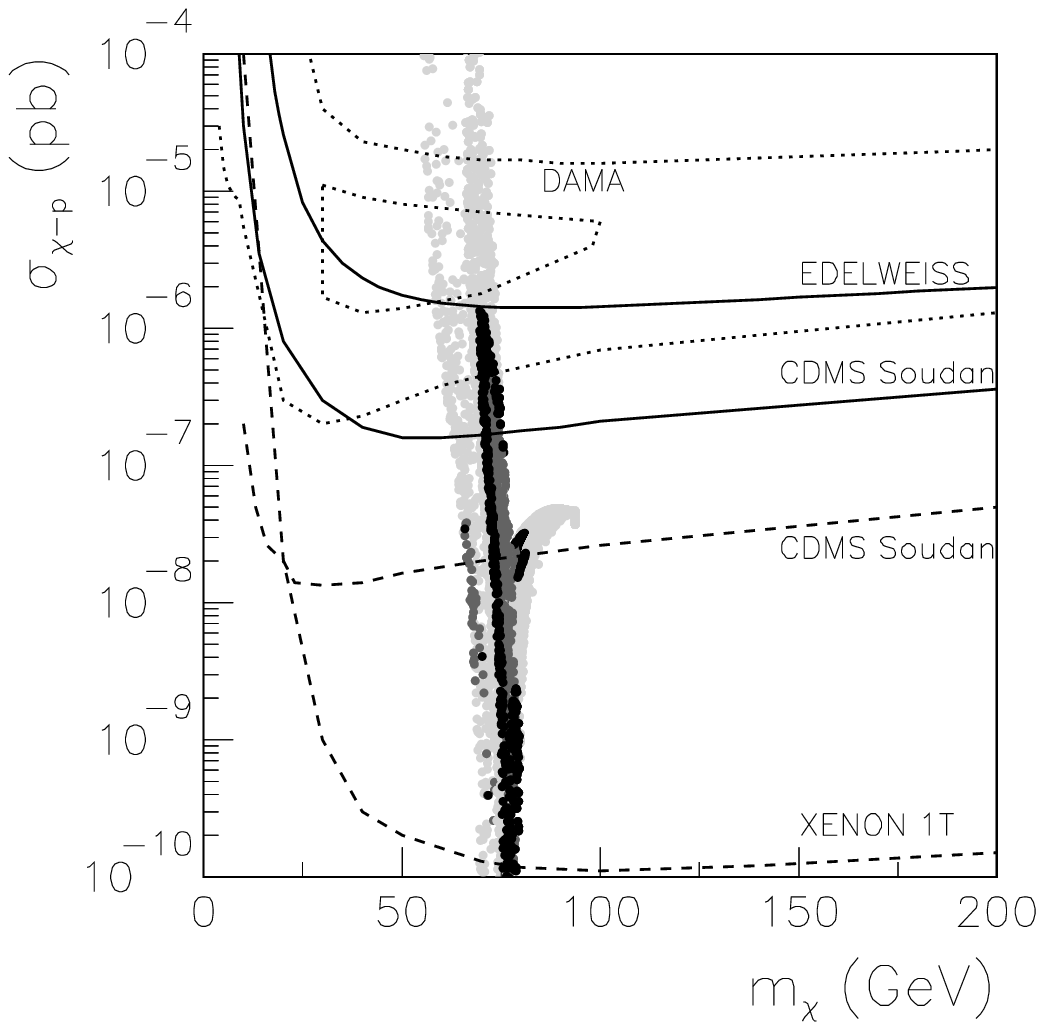}
\end{center}
\caption{Neutralino phenomenology in the NMSSM scenario. {\bf Left:}
Contour plot of the neutralino relic abundance as a function of the
superpotential parameters $\lambda$ and $k$. {\bf Right:}
Neutralino--nucleon cross section as a function of the neutralino
mass. Both figures are taken from \protect\cite{nmssm}.
\label{fig:NMSSM}}
\end{figure}

The superpotential of the NMSSM \cite{nmssm} contains one additional
Higgs singlet $S$ compared to the SUGRA or the MSSM scenarios: $
W=W_{MSSM}-\epsilon_{ij}\lambda S H_1^{i} H_2^{j}+1/3 k S^3$.  Its vev
generates dynamically the $\mu$ parameter at the EW scale, explaining why
the natural scale of
$\mu$, which is the only dimensional parameter in the superpotential,
is expected on general grounds to be at the GUT or Planck scale, while
it is required to be at the EW scale by phenomenology (the so--called
``$\mu$ problem'' of the MSSM). The main two features of the NSSMS are that one
Higgs scalar can be very light because it evades experimental limits
by mixing with its singlet component, and that the neutralino has an
additional singlino component, $ \chi\equiv a_1 \tilde{B}+a_2
\tilde{W_3}+a_3 \tilde{H}_1 +a_4 \tilde{H}_2+a_5 \tilde{S}$. An example
of relic neutralino phenomenology in the NMSSM is shown in
Fig.\ref{fig:NMSSM}, where on the left-hand side the cosmologically
allowed regions are shown in gray in the $\lambda$ and $k$ plane.  In
this scenario the presence of additional light Higgs bosons implies
more decay channels and resonant decays for the neutralino, which
tends to be relatively light and mostly singlino.  Thanks to the light
Higgs boson exchanged in the propagator the direct detection cross
section can be strongly enhanced, as can be seen from the
right--hand panel in Fig.\ref{fig:NMSSM}.

\begin{figure}
\begin{center}
\hspace{-1cm}
\includegraphics*[height=7cm]{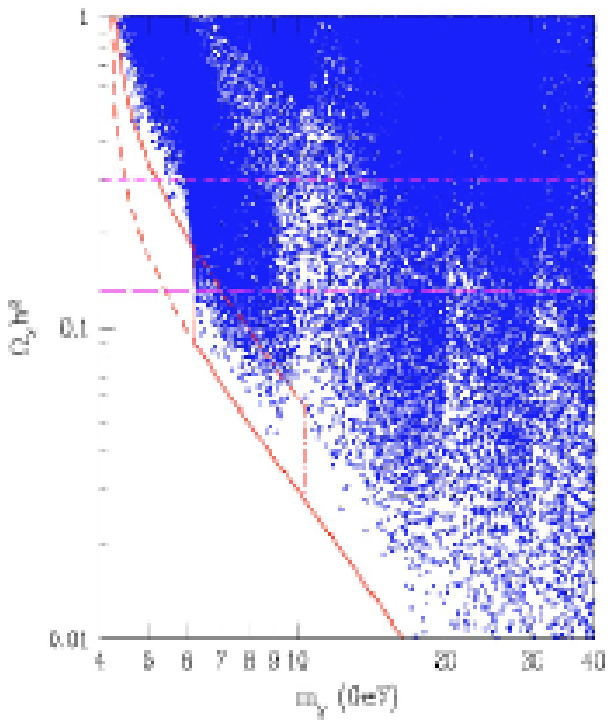}
\includegraphics*[height=7cm]{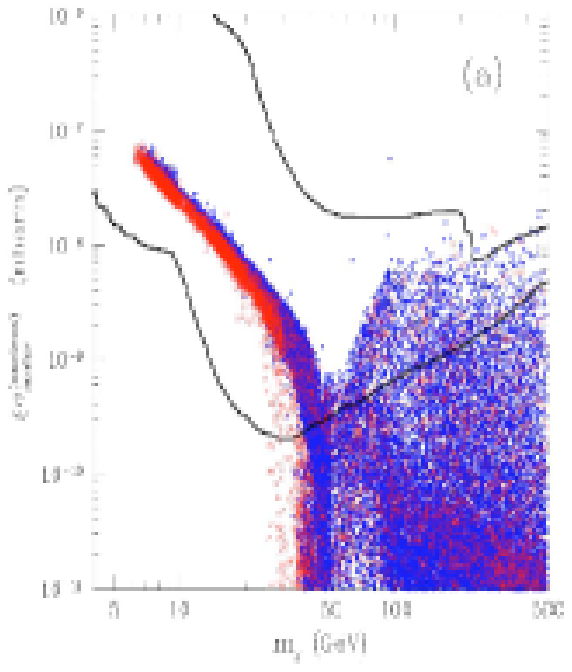}
\end{center}
\caption{ Neutralino phenomenology in an effective MSSM scenario with
$M_1<<M_2$.  {\bf Left:} Neutralino relic abundance as a function of
the neutralino mass\protect\cite{light_neutralinos}. {\bf Right:}
Neutralino--nucleon cross section as a function of the neutralino mass
\protect\cite{light_neutralinos_diretta}.
\label{fig:light_neutralinos}}
\end{figure}

In the effective MSSM all soft parameters are defined at the EW
scale. Usually in this scenario gaugino mass parameters are assumed to
be linked by the relation $M_1\simeq M_2/2$, which originates from
embedding the model in a GUT theory and by assuming that $M_1$ and
$M_2$ unify at the GUT scale. This relation implies the bound
$m_{\chi}\gsim m_{\chi^{\pm}}/2$, where the chargino mass
$m_{\chi^{\pm}}$ has been constrained by accelerator searches to be
heavier than about 100 GeV, so that $m_{\chi}\gsim 50$ GeV. The latter
limit, which is usually quoted for the neutralino mass, can however be
evaded if $M_1<<M_2$. In fact LSP searches at accelerators through the
production and decay of heavier neutralinos or the measurement of the
invisible width of the $Z$ boson are not sensitive enough to put a
direct lower bond on $m_{\chi}$. In this case, when standard
assumptions are made about the history of the Early Universe, a lower
bound $m_{\chi}\gsim$ 6 GeV can be derived by making use of the relic
abundance \cite{light_neutralinos}, as shown in the left panel of
Fig.\ref{fig:light_neutralinos}. Moreover, since $\Omega_{\chi}$ and
$\sigma^{\rm (nucleon)}_{\rm scalar}$ are strongly anticorrelated, an
upper bound on the former implies a lower bound on the latter.  This
explains why the cross section $\sigma^{\rm (nucleon)}_{\rm scalar}$
shown in the right--hand panel of Fig.  \ref{fig:light_neutralinos}
presents a typical low--mass funnel
\cite{light_neutralinos_diretta}. The theoretical points overlap
nicely with the region enclosed by the solid black line, which is
compatible to the annual modulation effect observed by the DAMA
experiment\cite{dama}.  This region is appropriately enlarged in order
to take into account the astrophysical uncertainties on the WIMP
velocity distribution function as well as the WIMP local density that
can strongly affect rate predictions \cite{modeling}, especially at
low WIMP masses.  This is due to the fact that in direct searches
lighter WIMPs need to kick a nucleus above the experimental recoil
energy threshold with a higher incoming velocity, so that very light
WIMPs probe the high--velocity tail of the distribution, which is more
sensitive to the details of the halo modeling. Note, however, that
exclusion plots calculated by experimental collaborations are usually
obtained by making use of a simplified isothermal sphere model for the
velocity distribution, so that some care is required in order to
assess in a consistent way how they affect low--mass WIMPs
\cite{uncertainties_cdms}. Moreover, recently the modulation effect
was re--analyzed by the DAMA Collaboration in light of a possible
effect of channeling in the NaI crystals \cite{damalast}. This effect
may lead to a significant enhancement of the detector response at low
recoil energies, displacing the DAMA region to lower WIMP masses, and
with direct consequences for the phenomenology of the susy
configurations that can explain the annual modulation \cite{noilast}.

\section{Conclusions}
WIMPs at the TeV scale can be realized in different well--motivated
scenarios, like the KK--photon in UED, the heavy photon in Little
Higgs and the neutralino in Supersymmetry. All these scenarios can
provide CDM with the correct abundance. Neutralino is still the most
popular, and today is available in different flavours, including
SUGRA, nuSUGRA, sub--GUT, Mirage mediation, NMSSM, effective MSSM,
scenarios with CP violation. Interestingly enough, some of these
scenarios are already at the level of present sensitivities for direct
DM searches.

\end{document}